\documentstyle[12pt,fleqn]{article}

\def\eps{\varepsilon}

\def\eft{\tilde{E}_{\rm f}}
\def\ef{E_{\rm f}}
\def\tk{T_{\rm K}}
\def\vt{\tilde{V}}
\def\beeq{\begin{equation}}
\def\eneq{\end{equation}}
\def\beeqa{\begin{eqnarray}}
\def\eneqa{\end{eqnarray}}

\def\eps{\varepsilon}

\addtocounter{section}{-1}
\begin{document}

\begin{center}

{\large {\bf Elastic anomaly of heavy fermion systems\\
in a crystalline field}\\
}
{Short title: {\sl Elastic anomaly of heavy fermion systems}}\\

\vspace{1cm}

{\rm Kikuo Harigaya\footnote[2]{Permanent address:
Fundamental Physics Section, Physical Science Division,
Electrotechnical Laboratory,
Umezono 1-1-4, Tsukuba, Ibaraki 305, Japan;
E-mail address: harigaya@etl.go.jp.}
and G. A. Gehring
}\\

\vspace{1cm}

{\sl Department of Physics, University of Sheffield,\\
Sheffield S3 7RH, United Kingdom}\\

\vspace{1cm}

(Received May 17, 1993)
\end{center}

\vspace{1cm}

\noindent
{\bf Abstract}\\
An elastic anomaly, observed in the heavy fermi liquid state of Ce
alloys (for example, CeCu$_6$ and CeTe), is analyzed by using the
infinite-$U$ Anderson lattice model.  The four atomic energy levels
are assumed for f-electrons.  Two of them are mutually degenerate.
A small crystalline splitting $2\Delta$ is assumed between two
energy levels.  The fourfold degenerate conduction bands are also
considered in the model.  We solve the model using the mean field
approximation to slave bosons, changing the Fermi energy in order
to keep the total electron number constant.  The nonzero value of
the mean field of the slave bosons persists over the temperatures
much higher than the Kondo temperature.  This is the effect of the
constant electron number.  Next, the linear susceptibility with
respect to $\Delta$ is calculated in order to obtain the renomalized
elastic constant.  The resulting temperature dependence of the
constant shows the downward dip.  We point out the relation of our
finding with the experimental data.

{}~

\noindent

\pagebreak


\section{Introduction}

Generally speaking, the atomic f-level of heavy fermion Ce
compounds is split by a crystalline field into multi sublevels [1].
The original f-level of the degeneracy 14 is first split into the
$j=5/2$ and $7/2$ levels by the $l$-$s$ interaction.  The $j=7/2$
level is about $10^3$K higher than the $j=5/2$ level, so the
main contribution comes from the $j = 5/2$ level, and the $j = 7/2$
level can be neglected.  Secondly, the $j=5/2$ level splits into
several multiplets in the presence of the crystalline field.
For example, in the cubic field, the $j=5/2$ level is composed
of two sublevels of degeneracy two and four, and in the tetragonal
field it splits into three Kramers doublets.  When the ground level
is doubly degenerate, the lowest f-level is nearly half filled
because the number of the 4f-electron is close to 0.9.  The
splitting width is normally $10^{1-2}$K, so the temperature
dependence of the physical quantities reflects the effect of
the thermal excitations among the sublevels.

The effect of the crystalline field on the single site Kondo system
has been one of recent topics in the theory of dilute magnetic alloys.
The Anderson model or the Coqblin-Schrieffer model, with the inclusion
of the crystalline field and the anisotropy of the mixing interaction,
is solved by the use of the Bethe ansatz technique [2] and the
self-consistent renomalization [3].  The result shows the variation
of the Kondo temperature $\tk$ with the temperature.  It comes from the
change of the impurity scattering channel.  For example, we
imagine the case where there are three Kramers doublets with the
splitting $\Delta_1$ amd $\Delta_2$.  In the temperature range
$T < \Delta_1$, the three levels act independently in the scattering.
But, in $\Delta_1 < T < \Delta_2$, the lower two levels behave
as if they are one level of the degeneracy four, because thermal
fluctuation has a larger width than $\Delta_1$.  In the same way,
in $T>\Delta_2$, all the three levels look as a level sixfold degenerate.
It has been discussed that the above features are exhibited in
temperature dependence of the magnetic susceptibility.

Experimentally, the heavy fermion systems show elastic anomalies
related with valence instabilities in low temperatures.  The most
striking effect is the softening of the elastic constants below the
Kondo temperature which are observed prominently in CeAl$_3$ [4],
CeCu$_6$ [5], and CeRu$_2$Si$_2$ [6].  The temperature dependences
of the elastic constants have been theoretically fitted quantitatively
well [7] by using the electron phonon coupling derived from the ansatz
that the mixing strength depends on the volume of the crystal.
The same ansatz was highly successful in the theoretical description
[8,9] of the Kondo volume collapse transition between $\alpha$-
and $\gamma$-Ce.  These elastic anomalies are related with the
overall instability of the valence mainly due to the change of
the mixing interaction.  They are observed in the wide temperature
range $0 < T < \tk$ and rather insensitive to the detailed
structures of the electronic band structures.

However, when we look at detailed temperature dependences of the
elastic constants, several compounds show anomalies which have
been interpreted as the crystalline field effect.  For example,
the elastic constant $C_{33}$ of CeCu$_6$ [10] has a dip at
about 10K.  This might be due to the splitting larger than
the Kondo temperature 4K.  The constant $(C_{11}-C_{12})/2$
of CeTe [11] shows the apparent dip at about 15K.  The ground
state levels of Ce ions split into the $\Gamma_7$ Kramers
doublet and the $\Gamma_8$ quartet states.  The $\Gamma_7$
states are the ground states.  There is the splitting 30K
between $\Gamma_7$ and $\Gamma_8$ states.  This is the
origin of the dip.

The main purpose of this paper is to present a microscopic
calculation in order to look at how the degeneracy structure and
crystalline field appear in the elastic properties.  The formalism
is independent of the real band structures and degeneracy structures.
Therefore, the results should be general enough for heavy fermion
systems.  We simply assume four quantum numbers for f-electrons.
Two of them have the same atomic energy level.  Thus, two different
atomic levels are assumed.  The difference of them, smaller than $\tk$,
means the width of the crystalline field splitting.
And, four conduction bands are assumed.  The total electronic
system is described by the infinite-$U$ Anderson lattice hamiltonian
by the slave boson method.  The model is solved by the mean field
approximation.  The formalism is explained in \S 2.

Firstly, we show mean field solutions and see that the mean field
value of slave bosons does not vanish  even at $T > \tk$.
This interesting property has been discussed previously [12,13].
It is owing to the fact that the Fermi level is varied in order to
keep the total electron number constant.  This is shown in \S 3.

Secondly, we calculate the linear susceptibility with respect to the
crystalline field splitting.  At lower temperatures, the susceptibility
has a structure related with the crystalline field effect.
The high temperature susceptibility does not depend on the splitting
width due to the large thermal excitation.  There is an upward dip in the
temperature dependence.  The dip is originated from the high degeneracy,
and its position moves with the splitting width.

Next, we assume an empirical relation between an elastic constant and the
susceptibility.  The form of the relation is assumed by taking account of
the quadrapoler response theory [14].  As the coupling constant between
the lattice and the crystalline field is unknown,
we should treat it as a kind of a parameter.  We show temperature
dependences of elastic constant for several choices of the coupling.
We will discuss relevant parameters for the elastic anomaly, i.e.,
the downward dip which is observed in the constant
$(C_{11}-C_{12})/2$ of CeTe [11].  The susceptibility and elastic
constant are reported and discussed in \S 4.

Finally, the paper is closed with several remarks in \S 5.

\section{Formalism}

We consider the infinite-$U$ Anderson lattice model in the
slave boson method.  The model has the following form:
\beeqa
H &=& \sum_{i}
[ ( E_{\rm f} - \Delta ) \sum_{l=1,2} f_{i,l}^\dagger f_{i,l}
+ ( E_{\rm f} + \Delta ) \sum_{l=3,4} f_{i,l}^\dagger f_{i,l} ] \\ \nonumber
&+& \sum_{{\vec k},l=1-4} \eps_{\vec k} c_{{\vec k},l}^\dagger
c_{{\vec k},l} \\ \nonumber
&+& V \sum_{i,l=1-4} ( f_{i,l}^\dagger c_{i,l} b_i
+ b_i^\dagger c_{i,l}^\dagger f_{i,l} ) \\ \nonumber
&+& \sum_i \lambda_i ( \sum_{l=1-4} f_{i,l}^\dagger f_{i,l}
+ b_i^\dagger b_i - 1),
\eneqa
where $f_{i,l}$ is an annihilation operator of the f-electron
of the $l$-th orbital at the $i$-th site, $c_{{\vec k},l}$ is
an operator of the conduction electron with the wave number
${\vec k}$, and $b_i$ is an operator of the slave boson which
indicates the unoccupied state at the f-orbital.  The atomic
energy of the first and second orbitals of f-electrons is
$E_{\rm f} - \Delta$, and that of the third and fourth orbitals
is $E_{\rm f} + \Delta$.  The crystalline field splitting
$2 \Delta$ is assumed between atomic energies of the two groups
of f-electrons.  For the conduction electrons, the same quantum
number is assumed as that of the f-electrons.  We use the square
density of states, $\rho \equiv 1/ND$, which extends over the
energy region, $-D < \eps_{\vec k} < (N-1) D$, where $N = 4$
is the total number of quantum states.  This assumes that the
combination $N \rho V^2$, which appears in the $1/N$ expansion,
is independent of $N$.  Therefore, the mean field theory becomes
exact as $N \rightarrow \infty$.  The third term in the
hamiltonian is the mixing interaction between f- and
c-electrons, $V$ being the interaction strength.  The last
term limits the maximum number of f-electrons per site up
to unity.  This could be realized by the constraint
$\sum_{l=1-4} f_{i,l}^\dagger f_{i,l} + b_i^\dagger b_i = 1$
with the Langrange multiplier field $\lambda_i$.

This model is treated within the mean field approximation:
$\langle b_i \rangle = r$, $\langle b_i^\dagger b_i \rangle
= r^2$, and $\lambda_i = \lambda $ (a site independent real value).
These mean field parameters are determined by solving the
following coupled equations:\\
(1) the constraint condition,
\beeqa
& & \frac{1}{2D} \int {\rm d}E \frac{\vt^2}{(\eft - \Delta - E)^2} f(E-\mu)
\\ \nonumber
&+& \frac{1}{2D} \int {\rm d}E \frac{\vt^2}{(\eft + \Delta - E)^2} f(E-\mu)
+ r^2 = 1,
\eneqa
(2) the self-consistency condition for $r$,
\beeqa
& & \frac{1}{2D} \int {\rm d}E \frac{V^2}{E- \eft + \Delta} f(E- \mu)
\\ \nonumber
&+& \frac{1}{2D} \int {\rm d}E \frac{V^2}{E- \eft - \Delta} f(E- \mu)
+ \lambda = 0,
\eneqa
and (3) the conservation condition of electron number $n_{\rm el}$,
\beeqa
& & \frac{1}{2D} \int {\rm d}E
[1 + \frac{\vt^2}{(\eft - \Delta - E)^2}] f(E-\mu) \\ \nonumber
&+& \frac{1}{2D} \int {\rm d}E
[1 + \frac{\vt^2}{(\eft + \Delta - E)^2}] f(E-\mu)
= n_{\rm el},
\eneqa
where $f(x) = 1/[{\rm exp}(x/T) + 1]$ is the Fermi distribution
function, $\eft = \ef + \lambda$ is the effective f-level, and
$\vt = rV$ is the effective mixing interaction.  The integrations
are performed over all the energy region of the bands.  The three
equations are solved numerically for the variables, $r, \lambda$,
and the Fermi level $\mu$.  In addition, the values at $T = 0$
can be obtained analytically.

\section{Solution}

Equations (2), (3), and (4) are solved numerically
for the parameters $D = 5 \times 10^4$K, $V = 7500$K,
$E_{\rm f} = - 10^4$K, and $n_{\rm el} = 1.9$ as
the typical values.  The splitting parameter $\Delta$
is increased from 0 but the magnitude is taken below
the Kondo temperature $\tk = \eft - \mu$.

Figure 1 shows the temperature dependences of parameters.
Figures 1 (a), (b), and (c) show the variations of $\eft$,
$\tk$, and the number of f-electrons per site $n_{\rm f}$,
respectively.  The data for $\Delta = 0, 15$, and 30K are shown
by the dashed, thin, and thick lines, respectively.  This
convention applies to all the figures in this paper.  As the
temperature increases, the order parameter $r$ decreases, so
that $n_{\rm f} = 1 - r^2$ increases.  The quantity $r$ does
not vanish even though the temperature is much higher than
$\tk$ at $T=0$.  This is the effect of the change of the Fermi
level $\mu$ to keep the total electron number constant.  This
effect has been reported previously [12,13].  According to the increase
of $n_{\rm f}$, $\eft$ decreases, which means the reduced
itinerancy of f-electrons.  At high temperatures, $T \gg \Delta$,
the three curves becomes almost identical, owing to the large
temperature excitation across the split $2 \Delta$.   At low
temperatures, the excitation energy is limited by the smaller
distance from the Fermi level to the gap of the bands $l = 1, 2$.
This results in the increased value of $n_{\rm f}$ when the
crystalline field is switched on.

We can derive the analytical expressions of parameters at $T=0$.
Small crystalline field splitting is assumed for the derivation.
The results are
\beeq
T_{\rm K} (\Delta) \equiv \eft - \mu = \sqrt{T_{\rm K}^2 (0) + \Delta^2}
\eneq
and
\beeq
r^2 \simeq \frac{ D T_{\rm K}^2 (0)}{V^2 T_{\rm K} (\Delta)},
\eneq
where $T_{\rm K} (0) = D \exp [-D (\mu - E_{\rm f})/V^2]$ is
the Kondo temperature for $\Delta = 0$.  These expressions
accord with the numerical results that both $\tk$ and
$n_{\rm f}$ increase as $\Delta$ increases.

\section{Elastic anomaly in low temperatures}

We shall discuss the change of elastic property of heavy fermions
due to the crystalline field splitting in the low temperature
below $\tk$.  Generally, an elastic constant $c$ is related with
the linear susceptibility with respect to $\Delta$,
as shown by the formula [14]:
\beeq
c = \frac{c_0}{1 + g \chi_\Delta},
\eneq
where $c_0$ is the elastic constant of the system where there is
not interactions between the lattice and the electronic system,
and $g$ is the coupling constant between $\Delta$ and the strain
field.  We assume that $c_0$ is independent of the temperature.
The value of $g$ is unknown experimentally as well as theoretically.
In order to discuss the crystalline field effect on $c$, we treat
the factor $g$ as a kind of fitting parameters.  The quantity
$\chi_\Delta$ is calculated as the second order derivative of
the mean field free energy:
\beeqa
\chi_\Delta &=& - \frac{\partial^2 F}{\partial \Delta^2}\\ \nonumber
&=& \frac{1}{D} \int dE \frac{\vt^2}{(\eft - \Delta - E)^3} f(E-\mu)
\\ \nonumber
&+& \frac{1}{D} \int dE \frac{\vt^2}{(\eft + \Delta - E)^3} f(E-\mu),
\eneqa
where the $\Delta$ dependences of the band edges are neglected
in the derivatives because their effect is exponentially small.

Figure 2 (a) displays the temperature dependence of $\chi_\Delta$.
The inverse of $\chi_\Delta$ is shown in Fig. 2(b).  There is a
peak at the low temperature.  When $\Delta = 0$, $\chi_\Delta$ is
identical with the magnetic susceptibility.  The peak of the curve
for $\Delta = 0$ has been found in the previous paper [12], and is the
effect of the fourfold degeneracy.  When $\Delta$ is finite, the
peak moves to the lower temperature and the peak height becomes
taller.  This is the crystalline field effect.  The curve at high
temperatures is less affected by $\Delta$.  The inverse of the
susceptibility in high temperatures is almost linear against $T$,
showing the Pauli behavior.  The increase of $\chi_\Delta$ in
the temperatures below the peak is well explained by the
analytical expression:
\beeq
\chi_\Delta = \frac{\tk^2 (0) + 2 \Delta^2}{\tk (\Delta) \tk^2 (0)},
\eneq
at $T = 0$.

We show the temperature dependences of $c/c_0$, assuming several
values for $g$.  We plot two sets of curves for $g = 1$K and 1.5K,
in Figs. 3(a) and (b), respectively.  The elastic constant decreases
from much higher to lower temperatures than $\tk$.  This is the
effect of the valence fluctuation.  There is a downward dip
around 15-20 K, which is the result of the combination of the
fourfold orbital degeneracy and the crystalline field splitting.
The larger $\Delta$ gives rise to the larger dip.

We shall look at relations with actual compounds.  The elastic
constant $C_{33}$ of CeCu$_6$ [10] has a dip at about 10K.  This might
be due to $\Delta$ larger than the Kondo temperature 4K.  The constant
$(C_{11}-C_{12})/2$ of CeTe [11] shows the apparent dip at about 15K.
The ground state levels of Ce ions in CeTe split into the $\Gamma_7$
Kramers doublet and the $\Gamma_8$ quartet states.  The $\Gamma_7$
states are the ground states.  There is the splitting $2\Delta = 30$K
between $\Gamma_7$ and $\Gamma_8$ states.  Fig. 1 of ref. [11] shows
that the decrease of $(C_{11} - C_{12})/2$ constant from that of
the ideal system without the electron--crystalline-field coupling
is about 4\% of the magnitude.  This behavior is simulated well by
the thin curve in Fig. 3(b).  The value $g = 1.5$K gives us a useful
information on the coupling strength of heavy fermions against
the crystalline field in CeTe.

\section{Concluding remarks}

We have solved the mean field equations of the Anderson lattice
model with crystalline field splitting $2 \Delta$, and calculated
the linear susceptibility with respect to $\Delta$.  Next, we
have simulated the temperature dependence of the elastic constant
which is derived by the coupling of the electronic system to the
crystalline field.  We have discussed the relation with experiments.

The downward dip of $c$ exists even if there is no crystalline
field splitting, i.e., $\Delta = 0$.  The finite $\Delta$ results
in the motion of the dip to lower temperatures.  The magnitude
of the dip becomes larger at the same time.  The temperature where
the dip is present would be finally determined by the combination
of the high degeneracy of the 4f orbitals and the splitting width.

One of the interesting future problems would be the magnetic field
effects.  Further removal of the degeneracy by the magnetic field
will give rise to more structures in the elastic constants.
The theory of the effects would be a helpful information for
experimental studies.

{}~

\noindent
{\bf Acknowledgements}\\
The authors should like to thank Professor Goto for bringing the problem
to their attention and for sending them copies of his papers.
This work was done while one of the authors (K.H.) was staying at
the University of Sheffield.  He acknowledges the financial support
for the stay and the hospitality that he obtained from the University.

\pagebreak

\begin{flushleft}
{\bf References}
\end{flushleft}

\noindent
$[1]$ Yamada K, Yoshida K and Hanzawa K 1984 {\sl Prog. Theor. Phys.}
{\bf 71} 450\\
$[2]$ Desgranges H U and Rasul J W 1987 {\sl Phys. Rev. B} {\bf 36}
328.\\
$[3]$ Kashiba S, Maekawa S, Takahashi S and Tachiki M
1986 {\sl J. Phys. Soc. Jpn.} {\bf 55} 1341.\\
$[4]$ Niksch M, L\"{u}thi B and Andres K 1980
{\sl Phys. Rev. B} {\bf 22} 5774\\
$[5]$ Weber D, Yoshizawa M, Kouroudis I, L\"{u}thi B
and Walker E 1987 {\sl Europhys. Lett.} {\bf 3} 827\\
$[6]$ L\"{u}thi B and Yoshizawa M 1987 {\sl J. Magn. Magn. Mater.}
{\bf 63/64} 274.\\
$[7]$ Thalmeier P 1987 {\sl J. Phys. C: Solid State Phys.}
{\bf 20} 4449\\
$[8]$ Allen J W and Martin R M 1982 {\sl Phys. Rev. Lett.}
{\bf 49} 1106\\
$[9]$ Lavagna M, Lacroix C and Cyrot M 1983 {\sl J. Phys. F:
Met. Phys.} {\bf 13} 1007.\\
$[10]$ Goto T, Suzuki T, Ohe Y, Fujimura T, Sakutsume S,
Onuki Y and Komatsubara T 1988 {\sl J. Phys. Soc. Jpn.}
{\bf 57} 2612\\
$[11]$ Matsui H, Goto T, Tamaki A, Fujimura T, Suzuki T and
Kasuya T 1988 {\sl J. Magn. Magn. Mater.} {\bf 76/77} 321\\
$[12]$ Evans S M M, Chung T and Gehring G A 1989
{\sl J. Phys.: Condens. Matter} {\bf 1} 10473\\
$[13]$ Harigaya K 1988 {\sl Master Thesis}, University of Tokyo\\
$[14]$ Levy P 1973 {\sl J. Phys. C: Solid State Phys.}
{\bf 6} 3545\\

\pagebreak

\begin{flushleft}
{\bf Figure Captions}
\end{flushleft}

\noindent
Fig. 1.  Temperature dependences of the mean field solution:
(a) $\eft$, (b) $\tk$, and (c) $n_{\rm f}$.  Parameters are
$D = 5 \times 10^4$K, $V = 7500$K, $E_{\rm f} = - 10^4$K,
and $n_{\rm el} = 1.9$.  The dashed, thin, and heavy lines
are for $\Delta = 0$, 15, and 30K, respectively.

{}~

\noindent
Fig. 2.  Temperature dependences of the linear susceptibility:
(a) $\chi_\Delta$ and (b) $1/\chi_\Delta$.  The parameters
and notations are the same as in Fig. 1.

{}~

\noindent
Fig. 3.  Temperature dependences of the elastic constant $c/c_0$
for (a) $g = 1.0$K and (b) $g = 1.5$K.  The parameters
and notations are the same as in Fig. 1.

\end{document}